\documentstyle[12pt]{article}
\newcommand{\be}{\begin{equation}}
\newcommand{\ee}{\end{equation}}
\newcommand{\bea}{\begin{eqnarray}}
\newcommand{\eea}{\end{eqnarray}}
\newcommand{\un}{\underline}
\newcommand{\eps}{\varepsilon}
\newcommand{\zahlen}{{\rm Z \!\! Z}}
\newcommand{\half}{{\scriptstyle{{1\over 2}}}}
\newcommand{\ihalf}{{\scriptstyle{{1\over 2i}}}}

\newcommand{\third}{{\scriptstyle{{1\over 3}}}}

\newcommand{\sixtth}{{\scriptstyle{{1\over 16}}}}

\newcommand{\real}{\relax{\rm I\kern-.18em R}}
\newcommand{\integer}{\relax{\rm I\kern-.18em N}}
\newcommand{\Tr}{{\rm Tr}}
\newcommand{\RE}{{\rm Re~}}
\newcommand{\IM}{{\rm Im~}}
\newcommand{\ad}{{\rm ad}}

\newcommand{\cD}{{\cal D}}
\newcommand{\cN}{{\cal N}}

\newcommand{\cO}{{\cal O}}
\newcommand{\cE}{{\cal E}}

\newcommand{\basispl}{
   \put(-.5,-.5){\line(1,0){1}}
   \put(.5,-.5){\line(0,1){1}}
   \put(.5,.5){\line(-1,0){1}}
   \put(-.5,.5){\line(0,-1){1}}
                         }
\newcommand{\basisar}{
   \put(0,-.5){\vector(1,0){0}}
   \put(.5,0){\vector(0,1){0}}
   \put(0,.5){\vector(-1,0){0}}
   \put(-.5,0){\vector(0,-1){0}}
	              }

\newcommand{\plaq}{\setlength{\unitlength}{.5cm}\raisebox{-.2cm}{
   \begin{picture}(1.2,1.2)(-.6,-.6)
   \basispl\basisar
   \put(-.5,-.5){\circle*{.2}}
   \put(-.55,-.55){\makebox(0,0)[tr]{\footnotesize $x$}}
   \put(-.55,0){\makebox(0,0)[r]{\footnotesize $\nu$}}
   \put(0,-.55){\makebox(0,0)[t]{\footnotesize $\mu$}}
   \end{picture}}}
\newcommand{\twoplaq}{\setlength{\unitlength}{1cm}\raisebox{-.5cm}{
   \begin{picture}(1.2,1.2)(-.6,-.6)
   \basispl
   \put(-.5,-.5){\circle*{.1}}
   \put(-.5,.5){\circle*{.1}}
   \put(.5,-.5){\circle*{.1}}
   \put(.5,.5){\circle*{.1}}
   \put(0,-.5){\circle*{.1}}
   \put(0,.5){\circle*{.1}}
   \put(.5,0){\circle*{.1}}
   \put(-.5,0){\circle*{.1}}
   \put(-.25,-.5){\vector(1,0){0}}
   \put(.25,-.5){\vector(1,0){0}}
   \put(.5,-.25){\vector(0,1){0}}
   \put(.5,.25){\vector(0,1){0}}
   \put(-.25,.5){\vector(-1,0){0}}
   \put(.25,.5){\vector(-1,0){0}}
   \put(-.5,-.25){\vector(0,-1){0}}
   \put(-.5,.25){\vector(0,-1){0}}
   \put(-.55,-.55){\makebox(0,0)[tr]{\footnotesize $x$}}
   \put(-.55,0){\makebox(0,0)[r]{\footnotesize $\nu$}}
   \put(0,-.55){\makebox(0,0)[t]{\footnotesize $\mu$}}
   \end{picture}}}
\newcommand{\wplaqone}{\setlength{\unitlength}{1cm}\raisebox{-.5cm}{
   \begin{picture}(1.2,1.2)(-.6,-.6)
   \put(.25,-.5){\line(0,1){1}}
   \put(.25,.5){\line(-1,0){.5}}
   \put(-.25,.5){\line(0,-1){1}}
   \put(-.25,0){\line(1,0){.1}}
   \put(-.10,0){\line(1,0){.1}}
   \put(.05,0){\line(1,0){.1}}
   \put(.2,0){\line(1,0){.05}}
   \put(-.25,-.5){\circle*{.1}}
   \put(-.25,.5){\circle*{.1}}
   \put(.25,-.5){\circle*{.1}}
   \put(.25,.5){\circle*{.1}}
   \put(-.25,0){\circle*{.1}}
   \put(.25,0){\circle*{.1}}
   \put(-.25,-.15){\vector(0,1){0}}
   \put(-.25,.35){\vector(0,1){0}}
   \put(.1,.5){\vector(1,0){0}}
   \put(.25,.15){\vector(0,-1){0}}
   \put(.25,-.35){\vector(0,-1){0}}
   \put(-.3,-.55){\makebox(0,0)[tr]{\footnotesize $x$}}
   \put(-.34,-.3){\makebox(0,0)[r]{\footnotesize $a$}}
   \put(-.43,.3){\makebox(0,0)[t]{\footnotesize $b$}}
   \put(0,.8){\makebox(0,0)[t]{\footnotesize $\mu$}}
   \end{picture}}}
\newcommand{\wplaqonem}{\setlength{\unitlength}{1cm}\raisebox{-.5cm}{
   \begin{picture}(1.2,1.2)(-.6,-.6)
   \put(-.29,-.5){\line(0,1){.54}}
   \put(-.29,.04){\line(1,0){.54}}
   \put(.25,.04){\line(0,1){.46}}
   \put(.25,.5){\line(-1,0){.46}}
   \put(-.21,.5){\line(0,-1){.40}}
   \put(-.21,-.04){\line(1,0){.46}}
   \put(.25,-.04){\line(0,-1){.46}}
   \put(-.25,-.5){\circle*{.1}}
   \put(-.21,.5){\circle*{.1}}
   \put(.25,.5){\circle*{.1}}
   \put(-.29,-.5){\circle*{.1}}
   \put(.25,-.5){\circle*{.1}}
   \put(-.29,-.15){\vector(0,1){0}}
   \put(-.21,.25){\vector(0,-1){0}}
   \put(-.1,.5){\vector(-1,0){0}}
   \put(.2,-.04){\vector(1,0){0}}
   \put(.1,.04){\vector(1,0){0}}
   \put(.25,.35){\vector(0,1){0}}
   \put(.25,-.35){\vector(0,-1){0}}
   \put(-.3,-.55){\makebox(0,0)[tr]{\footnotesize $x$}}
   \put(-.34,-.3){\makebox(0,0)[r]{\footnotesize $a$}}
   \put(.4,.45){\makebox(0,0)[t]{\footnotesize $b$}}
   \put(0,.3){\makebox(0,0)[t]{\footnotesize $\mu$}}
   \end{picture}}}
\newcommand{\wplaqtwo}{\setlength{\unitlength}{1cm}\raisebox{-.5cm}{
   \begin{picture}(1.2,1.2)(-.6,-.6)
   \put(-.5,-.25){\line(0,1){.5}}
   \put(-.5,.25){\line(1,0){1}}
   \put(.5,.25){\line(0,-1){.5}}
   \put(.5,-.25){\line(-1,0){.5}}
   \put(0,-.25){\line(0,1){.1}}
   \put(0,-.10){\line(0,1){.1}}
   \put(0,.05){\line(0,1){.1}}
   \put(0,.20){\line(0,1){.1}}
   \put(-.5,-.25){\circle*{.1}}
   \put(0,-.25){\circle*{.1}}
   \put(.5,-.25){\circle*{.1}}
   \put(-.5,.25){\circle*{.1}}
   \put(0,.25){\circle*{.1}}
   \put(-.5,.25){\circle*{.1}}
   \put(-.5,.1){\vector(0,1){0}}
   \put(-.15,.25){\vector(1,0){0}}
   \put(.35,.25){\vector(1,0){0}}
   \put(.5,-.1){\vector(0,-1){0}}
   \put(.2,-.25){\vector(-1,0){0}}
   \put(-.55,-.3){\makebox(0,0)[tr]{\footnotesize $x$}}
   \put(.35,.42){\makebox(0,0)[r]{\footnotesize $a$}}
   \put(-.65,.15){\makebox(0,0)[t]{\footnotesize $b$}}
   \put(-.25,.5){\makebox(0,0)[t]{\footnotesize $\mu$}}
   \end{picture}}}
\newcommand{\wplaqtwom}{\setlength{\unitlength}{1cm}\raisebox{-.5cm}{
   \begin{picture}(1.2,1.2)(-.6,-.6)
   \put(-.5,-.25){\line(0,1){.54}}
   \put(-.5,.29){\line(1,0){.54}}
   \put(.5,.21){\line(-1,0){.40}}
   \put(.5,.21){\line(0,-1){.46}}
   \put(.5,-.25){\line(-1,0){.46}}
   \put(.04,.29){\line(0,-1){.54}}
   \put(-.04,.21){\line(0,-1){.46}}
   \put(-.5,-.25){\circle*{.1}}
   \put(.5,-.25){\circle*{.1}}
   \put(-.5,.29){\circle*{.1}}
   \put(.5,.21){\circle*{.1}}
   \put(-.5,.1){\vector(0,1){0}}
   \put(-.15,.29){\vector(1,0){0}}
   \put(.2,.21){\vector(-1,0){0}}
   \put(.04,-.05){\vector(0,-1){0}}
   \put(-.04,-.15){\vector(0,-1){0}}
   \put(.5,.1){\vector(0,1){0}}
   \put(.35,-.25){\vector(1,0){0}}
   \put(-.55,-.3){\makebox(0,0)[tr]{\footnotesize $x$}}
   \put(.35,-.42){\makebox(0,0)[r]{\footnotesize $a$}}
   \put(-.65,.15){\makebox(0,0)[t]{\footnotesize $b$}}
   \put(-.25,.54){\makebox(0,0)[t]{\footnotesize $\mu$}}
   \end{picture}}}
\newcommand{\wplaqthree}{\setlength{\unitlength}{1cm}\raisebox{-.5cm}{
   \begin{picture}(1.2,1.2)(-.6,-.6)
   \put(-.5,-.25){\line(0,1){.5}}
   \put(-.5,.25){\line(1,0){1}}
   \put(.5,.25){\line(0,-1){.5}}
   \put(0,-.25){\line(-1,0){.5}}
   \put(0,-.25){\line(0,1){.1}}
   \put(0,-.10){\line(0,1){.1}}
   \put(0,.05){\line(0,1){.1}}
   \put(0,.20){\line(0,1){.1}}
   \put(-.5,-.25){\circle*{.1}}
   \put(0,-.25){\circle*{.1}}
   \put(.5,-.25){\circle*{.1}}
   \put(-.5,.25){\circle*{.1}}
   \put(0,.25){\circle*{.1}}
   \put(-.5,.25){\circle*{.1}}
   \put(-.5,.1){\vector(0,1){0}}
   \put(.35,.25){\vector(1,0){0}}
   \put(-.15,.25){\vector(1,0){0}}
   \put(.5,-.05){\vector(0,-1){0}}
   \put(-.35,-.25){\vector(-1,0){0}}
   \put(.1,-.35){\makebox(0,0)[tr]{\footnotesize $x$}}
   \put(.37,.42){\makebox(0,0)[r]{\footnotesize $\mu$}}
   \put(-.65,.15){\makebox(0,0)[t]{\footnotesize $b$}}
   \put(-.25,.5){\makebox(0,0)[t]{\footnotesize $a$}}
   \end{picture}}}
\newcommand{\wplaqthreem}{\setlength{\unitlength}{1cm}\raisebox{-.5cm}{
   \begin{picture}(1.2,1.2)(-.6,-.6)
   \put(.5,-.25){\line(0,1){.54}}
   \put(.5,.29){\line(-1,0){.54}}
   \put(-.5,.21){\line(1,0){.40}}
   \put(-.5,.21){\line(0,-1){.46}}
   \put(-.5,-.25){\line(1,0){.46}}
   \put(-.04,.29){\line(0,-1){.54}}
   \put(.04,.21){\line(0,-1){.46}}
   \put(.5,-.25){\circle*{.1}}
   \put(-.5,-.25){\circle*{.1}}
   \put(.5,.29){\circle*{.1}}
   \put(-.5,.21){\circle*{.1}}
   \put(.5,-.1){\vector(0,-1){0}}
   \put(.35,.29){\vector(1,0){0}}
   \put(-.35,.21){\vector(-1,0){0}}
   \put(-.04,.05){\vector(0,1){0}}
   \put(.04,.2){\vector(0,1){0}}
   \put(-.5,-.1){\vector(0,-1){0}}
   \put(-.15,-.25){\vector(1,0){0}}
   \put(.15,-.35){\makebox(0,0)[tr]{\footnotesize $x$}}
   \put(-.2,-.42){\makebox(0,0)[r]{\footnotesize $a$}}
   \put(-.15,.15){\makebox(0,0)[t]{\footnotesize $b$}}
   \put(.25,.54){\makebox(0,0)[t]{\footnotesize $\mu$}}
   \end{picture}}}
\newcommand{\twoone}{\setlength{\unitlength}{1cm}\raisebox{-.5cm}{
   \begin{picture}(1.2,1.2)(-.6,-.6)
   \put(0,-.5){\line(1,0){.5}}
   \put(.5,-.5){\line(0,1){1}}
   \put(.5,.5){\line(-1,0){1}}
   \put(-.5,.5){\line(0,-1){1}}
   \put(-.5,-.5){\circle*{.1}}
   \put(-.5,.5){\circle*{.1}}
   \put(.5,-.5){\circle*{.1}}
   \put(.5,.5){\circle*{.1}}
   \put(0,-.5){\circle*{.1}}
   \put(0,.5){\circle*{.1}}
   \put(.5,0){\circle*{.1}}
   \put(-.5,0){\circle*{.1}}
   \put(.25,-.5){\vector(-1,0){0}}
   \put(.5,-.25){\vector(0,-1){0}}
   \put(.5,.25){\vector(0,-1){0}}
   \put(-.25,.5){\vector(1,0){0}}
   \put(.25,.5){\vector(1,0){0}}
   \put(-.5,-.25){\vector(0,1){0}}
   \put(-.5,.25){\vector(0,1){0}}
   \put(-.55,-.55){\makebox(0,0)[tr]{\footnotesize $x$}}
   \put(-.55,0){\makebox(0,0)[r]{\footnotesize $\nu$}}
   \put(.2,-.55){\makebox(0,0)[t]{\footnotesize $\mu$}}
   \end{picture}}}
\newcommand{\twotwo}{\setlength{\unitlength}{1cm}\raisebox{-.5cm}{
   \begin{picture}(1.2,1.2)(-.6,-.6)
   \put(-.5,-.5){\line(1,0){.5}}
   \put(.5,-.5){\line(0,1){1}}
   \put(.5,.5){\line(-1,0){1}}
   \put(-.5,.5){\line(0,-1){1}}
   \put(-.5,-.5){\circle*{.1}}
   \put(-.5,.5){\circle*{.1}}
   \put(.5,-.5){\circle*{.1}}
   \put(.5,.5){\circle*{.1}}
   \put(0,-.5){\circle*{.1}}
   \put(0,.5){\circle*{.1}}
   \put(.5,0){\circle*{.1}}
   \put(-.5,0){\circle*{.1}}
   \put(-.25,-.5){\vector(-1,0){0}}
   \put(.5,-.25){\vector(0,-1){0}}
   \put(.5,.25){\vector(0,-1){0}}
   \put(-.25,.5){\vector(1,0){0}}
   \put(.25,.5){\vector(1,0){0}}
   \put(-.5,-.25){\vector(0,1){0}}
   \put(-.5,.25){\vector(0,1){0}}
   \put(-.25,-.55){\makebox(0,0)[tr]{\footnotesize $\mu$}}
   \put(-.55,0){\makebox(0,0)[r]{\footnotesize $\nu$}}
   \put(0.1,-.55){\makebox(0,0)[t]{\footnotesize $x$}}
   \end{picture}}}
\newcommand{\twothree}{\setlength{\unitlength}{1cm}\raisebox{-.5cm}{
   \begin{picture}(1.2,1.2)(-.6,-.6)
   \put(-.5,-.5){\line(1,0){1}}
   \put(.5,-.5){\line(0,1){1}}
   \put(.5,.5){\line(-1,0){.5}}
   \put(-.5,.5){\line(0,-1){1}}
   \put(-.5,-.5){\circle*{.1}}
   \put(-.5,.5){\circle*{.1}}
   \put(.5,-.5){\circle*{.1}}
   \put(.5,.5){\circle*{.1}}
   \put(0,-.5){\circle*{.1}}
   \put(0,.5){\circle*{.1}}
   \put(.5,0){\circle*{.1}}
   \put(-.5,0){\circle*{.1}}
   \put(-.25,-.5){\vector(1,0){0}}
   \put(.25,-.5){\vector(1,0){0}}
   \put(.5,-.25){\vector(0,1){0}}
   \put(.5,.25){\vector(0,1){0}}
   \put(.25,.5){\vector(-1,0){0}}
   \put(-.5,-.25){\vector(0,-1){0}}
   \put(-.5,.25){\vector(0,-1){0}}
   \put(-.5,.7){\makebox(0,0)[tr]{\footnotesize $x$}}
   \put(-.55,0){\makebox(0,0)[r]{\footnotesize $\nu$}}
   \put(0.3,.7){\makebox(0,0)[t]{\footnotesize $\mu$}}
   \end{picture}}}
\newcommand{\twofour}{\setlength{\unitlength}{1cm}\raisebox{-.5cm}{
   \begin{picture}(1.2,1.2)(-.6,-.6)
   \put(-.5,-.5){\line(1,0){1}}
   \put(.5,-.5){\line(0,1){1}}
   \put(.0,.5){\line(-1,0){.5}}
   \put(-.5,.5){\line(0,-1){1}}
   \put(-.5,-.5){\circle*{.1}}
   \put(-.5,.5){\circle*{.1}}
   \put(.5,-.5){\circle*{.1}}
   \put(.5,.5){\circle*{.1}}
   \put(0,-.5){\circle*{.1}}
   \put(0,.5){\circle*{.1}}
   \put(.5,0){\circle*{.1}}
   \put(-.5,0){\circle*{.1}}
   \put(-.25,-.5){\vector(1,0){0}}
   \put(.25,-.5){\vector(1,0){0}}
   \put(.5,-.25){\vector(0,1){0}}
   \put(.5,.25){\vector(0,1){0}}
   \put(-.25,.5){\vector(-1,0){0}}
   \put(-.5,-.25){\vector(0,-1){0}}
   \put(-.5,.25){\vector(0,-1){0}}
   \put(0.25,.7){\makebox(0,0)[tr]{\footnotesize $x$}}
   \put(-.55,0){\makebox(0,0)[r]{\footnotesize $\nu$}}
   \put(-.25,.7){\makebox(0,0)[t]{\footnotesize $\mu$}}
   \end{picture}}}
\newcommand{\stapup}{\setlength{\unitlength}{.5cm}\raisebox{-.2cm}{
   \begin{picture}(1.2,1.2)(-.6,-.6)
   \put(.5,-.5){\line(0,1){1}}
   \put(.5,.5){\line(-1,0){1}}
   \put(-.5,.5){\line(0,-1){1}}
   \put(.5,0){\vector(0,-1){0}}
   \put(0,.5){\vector(1,0){0}}
   \put(-.5,0){\vector(0,1){0}}
   \put(-.5,-.5){\circle*{.2}}
   \put(-.55,-.55){\makebox(0,0)[tr]{\footnotesize $x$}}
   \put(-.55,0){\makebox(0,0)[r]{\footnotesize $\nu$}}
   \put(0,.55){\makebox(0,0)[b]{\footnotesize $\mu$}}
   \end{picture}}}
\newcommand{\stapdw}{\setlength{\unitlength}{.5cm}\raisebox{-.2cm}{
   \begin{picture}(1.2,1.2)(-.6,-.6)
   \put(.5,-.5){\line(0,1){1}}
   \put(.5,-.5){\line(-1,0){1}}
   \put(-.5,.5){\line(0,-1){1}}
   \put(.5,0){\vector(0,1){0}}
   \put(0,-.5){\vector(1,0){0}}
   \put(-.5,0){\vector(0,-1){0}}
   \put(-.5,.5){\circle*{.2}}
   \put(-.55,.75){\makebox(0,0)[tr]{\footnotesize $x$}}
   \put(-.55,0){\makebox(0,0)[r]{\footnotesize $\nu$}}
   \put(0,-.55){\makebox(0,0)[t]{\footnotesize $\mu$}}
   \end{picture}}}

\def\phm{\hphantom{-}}
\def\pho{\hphantom{1}}
\def\mystrut{{\vrule height 15pt depth 4pt width 0pt}}

\begin{document}
\vskip-1cm
\hfill INLO-PUB-2/94
\vskip5mm
\begin{center}
{\LARGE{\bf{\underline{Sphalerons and other saddles from cooling}}}}\\
\vspace*{1cm}{\large
Margarita Garc\'{\i}a P\'erez and Pierre van Baal\\}
\vspace*{1cm}
Instituut-Lorentz for Theoretical Physics,\\
University of Leiden, PO Box 9506,\\
NL-2300 RA Leiden, The Netherlands.\\ 
\end{center}
\vspace*{5mm}{\narrower\narrower{\noindent
\underline{Abstract:} We describe a new cooling algorithm for SU(2) lattice
gauge theory. It has any critical point of the energy or action functional
as a fixed point. In particular, any number of unstable modes may occur.
We also provide insight in the convergence of the cooling algorithms.
A number of solutions will be discussed, in particular the sphalerons for
twisted and periodic boundary conditions which are important for the 
low-energy dynamics of gauge theories. For a unit cubic volume we find
a sphaleron energy of resp. $\cE_s=34.148(2)$ and $\cE_s=72.605(2)$ for the 
twisted and periodic case. Remarkably, the magnetic field for the periodic
sphaleron satisfies at all points $\Tr B_x^2=\Tr B_y^2=\Tr B_z^2$.
}\par}

\section{Introduction}

Saddle points of the energy functional can play an important role in 
non-perturbative dynamics, as they describe the minimal barrier to be taken to 
go from one to the next minimum energy configuration. In a heat bath it
determines the critical temperature where barrier crossing is no longer
suppressed.  In a strongly interacting theory it determines the coupling
(or volume) where barrier crossing - quantum mechanically always allowed - 
is no longer exponentially suppressed. The saddle point we have in mind here
is the sphaleron~\cite{kli}, which has precisely one unstable mode. This comes
about since one determines the minimal barrier height from a mini-max 
procedure~\cite{tau}. When one has a path in field space from one minimum
to the other, one first finds the maximal energy along the path. Then
one minimizes this maximum over the space of all such paths. The unstable
mode will correspond to the decrease along the path that goes through
this mini-max configuration. Given such a saddle point, it is often difficult
to prove in all rigour that it is the one with lowest energy, as there 
may be local minima when varying over the space of paths.

Quite often a sphaleron lies on a particular instanton path (i.e. the 
action associated to the path is also minimal), but this is by no means 
guaranteed. When the instanton has non-trivial parameters (moduli),
mini-max with respect to those parameters provides a candidate sphaleron.
For $S^3$ one can analytically verify that this candidate is a saddle
point of the energy functional with one unstable mode~\cite{baa1}.
For $T^3$ this was studied numerically~\cite{gar1,gar2}. It was not entirely
conclusive from that study if the sphaleron was indeed associated to 
the top of an instanton path. With the new algorithm we have been able
to address this issue more directly and conclude that it is the case
for both periodic and twisted~\cite{gar3} boundary conditions. 

We will only consider the case of pure SU(2) gauge theories. In an infinite
volume the mini-max procedure is easily seen to lead to zero sphaleron
energy, as instantons can have an arbitrarily large scale parameter.
In a finite volume, the scale parameter can not be larger than the 
volume, which stabilizes the mini-max procedure at a non-zero energy.
In the presence of a Higgs field, like in the standard model~\cite{kli},
the scale is set by the Higgs mass.

Before discussing the numerical results we will first describe the 
algorithm for finding saddle points and discuss the convergence.
This study is specific for SU(2), but the principle of taking the square
of the equations of motion as the action to be minimized will work equally
well for SU(N) gauge theories, with or without scalar fields. In section 4 we
will discuss results for the class of analytically known solutions with
constant field strength, to test the various aspects of the method.
In section 5 the results for the sphalerons are presented in detail.
We end the paper with some concluding remarks. Readers only interested
in the results should directly go to section 5, perhaps after section 2.

\section{The algorithm}

Consider the square of the equations of motion summed over its 
variables as a functional. Its absolute minima are by construction 
solutions of the equations of motion. Stability is now obvious, but 
also follows from the fact that at such a solution the Hessian of the 
new functional is (twice) the square of the Hessian of the energy (or action) 
functional. In the $n$-dimensional continuum we have the following expressions
for the two functionals
\be
S=-\half\int d_nx~\Tr(F_{\mu\nu}^2(x))\quad,\label{eq:Scon}
\ee
\be
\hat S=-2\int d_nx~\Tr((\cD_\mu F_{\mu\nu}(x))^2)\quad\label{eq:hScon},
\ee
with $F_{\mu\nu}(x)=\partial_\mu A_\nu(x)-\partial_\nu A_\mu(x) +[A_\mu(x),
A_\nu(x)]$ the Yang-Mills field strength in terms of the anti-hermitian
vector potential $A_\mu(x)$ and $\cD_\mu=\partial_\mu+\ad A_\mu(x)$ the 
covariant derivative.

This was considered in the past for constructing monopole solutions 
in three dimensions~\cite{dun}, but the process of finding the minimum
of $\hat S$ was not gauge invariant. Like for the Wilson action~\cite{wil} 
\be
S=N^{4-n}\sum_{x,\mu,\nu} \Tr\left(1-\plaq\right)=N^{4-n}\sum_{x,\mu,\nu} 
\Tr\left(1-U_\mu(x)U_\nu(x+\hat{\mu})U_\mu^{\dagger}(x+\hat{\nu})
U_\nu^{\dagger}(x)\right)\quad,\label{eq:Swil}
\ee
a gauge invariant lattice action for the square of the equations of motion 
is not too difficult to write down~\cite{sij}
\be
\hat S=N^{6-n}\sum_{x,\mu} 
\RE\Tr(\tilde{U}_\mu(x)\tilde{U}_\mu^{\dagger}(x)-(U_\mu(x)
\tilde{U}_\mu^{\dagger}(x))^2)\quad,\label{eq:Sarj}
\ee
where $U_\mu(x)$ is a group element on the link that runs from $x$ to
$x+\hat{\mu}$ ($\hat\mu$ is the unit vector in the $\mu$ direction),
$N$ is the number of lattice points in one of the directions
and $\tilde{U}_\mu(x)$ is the sum over $2(n-1)$ staples
\be
\tilde{U}_\mu(x)=\sum_{\nu\neq\mu}\left(\stapup+\stapdw\right)=
\sum_{\nu\neq\mu}(U_\nu(x)U_\mu(x+\hat{\nu})U_\nu^{\dagger}(x+\hat{\mu})  
+U_\nu^{\dagger}(x-\hat{\nu})U_\mu(x-\hat{\nu})U_\nu(x+\hat{\mu}-\hat{\nu}))
,\label{eq:Util} \ee
To derive the lattice equations of motion, we observe that $S$ depends on 
$U_\mu(x)$ through the expression:
\be
S(U_\mu(x))=2N^{4-n}\RE\Tr(1-U_\mu(x)\tilde{U}_\mu^{\dagger}(x))
\quad.\label{eq:Ssin}
\ee
As $\tilde{U}_\mu(x)$ is independent of $U_\mu(x)$, 
it follows that
\be
U_\mu(x)\tilde{U}_\mu^{\dagger}(x)-\tilde{U}_\mu(x)
U_\mu^{\dagger}(x)=0\quad.\label{eq:leqm}
\ee
Its square, summed over the lattice, yields $\hat S$. For SU(2) one can
also use the fact that the equations of motion are solved by
\be
U_\mu(x)=\pm\tilde{U}_\mu(x)/\|\tilde{U}_\mu(x)\|\quad,\quad
\|\tilde{U}_\mu(x)\|^2\equiv \half\Tr(\tilde{U}_\mu(x)\tilde{U}_\mu^{\dagger}
(x))\quad,\label{eq:Ssol} \ee
where only the positive sign will allow for solutions that have a smooth 
continuum limit. We can now introduce an error functional~\cite{gar2} 
\be
\tilde{S}=N^{6-n}\sum_{x,\mu}\Tr(1-U_\mu(x)\tilde{U}_\mu^{\dagger}
(x)/\|\tilde{U}_\mu(x)\|)\quad.\label{eq:Sus}
\ee
In the continuum limit one finds $\tilde{S}=\sixtth(n-1)^{-2}\hat S$.
This follows in the usual way~\cite{gar1}
from expanding the links in $a=1/N$, defining the relation between
the vector potential in the continuum and the link variables on the 
lattice by
\be
U_\mu(x)={\rm Pexp}(\int_0^a A_\mu(x+s\hat{\mu})ds)\quad.\label{eq:link}
\ee
One finds after some algebra 
\be
U_\mu(x)\tilde{U}_\mu^{\dagger}(x)/\|\tilde{U}_\mu(x)\|=\exp\left(-{1\over
{2(n-1)N^3}}\sum_\nu\cD_\nu F_{\mu\nu}(x)+\cO(N^{-5})\right)\quad.\label{eq:exp}
\ee

We can instead of the standard Wilson action use any improved action, containing
$n\times m$ Wilson loops with appropriate couplings, such as the 
over-improved actions we have used in ref.~\cite{gar1} 
\be
S(\eps)=\frac{4-\eps}{3}\sum_{x,\mu,\nu}\Tr\left(1-\plaq\right)
+\frac{\eps-1}{48}
\sum_{x,\mu,\nu}\Tr \left(1-\twoplaq\right)\quad.\label{eq:Seps}
\ee
All that needs to be modified in the above is the definition of 
$\tilde U_\mu(x)$, which will now also contain the staples associated
with the larger Wilson loops appearing in eq.~(\ref{eq:Seps})
\be
\tilde{U}_\mu(x)=\sum_{\nu\neq\mu}{4-\eps\over 3}\left(\stapup+\stapdw\right)+
{\eps-1\over 48}\left(\twoone+\twotwo+\twothree+\twofour\right)\quad.
\label{eq:stap2}
\ee
Like for $S(\eps)$, the $\eps$ dependence drops out in the continuum limit 
for $\hat S(\eps)$. However, for $\tilde S$ one finds in this limit
$\tilde S(\eps)=36((n-1)(31-7\eps))^{-2}\hat S$.  

The ordinary cooling algorithm~\cite{ber} for both of these actions is 
based on the observation that the replacement
\be
U_\mu^\prime(x)=\tilde{U}_\mu(x)/\|\tilde{U}_\mu(x)\|\quad,\label{eq:slup}
\ee
minimizes $S(U_\mu(x))$. Note that for $S$ (resp. $S(\eps)$) cooling we should 
require the lattice to be at least two (resp. three) sites in each direction. 
To completely specify the algorithm one also should prescribe in what order 
every link is being updated. One sweep is the process of updating each link 
precisely once in that particular order. This defines a mapping 
($U\equiv\{U_\mu(x)\}$)
\be
U^\prime=T(U)\quad,\quad T=T_\cN\circ T_{\cN-1}\circ\cdots\circ T_2\circ T_1
\quad,\label{eq:swee}
\ee
where $T_i$ is the single-link update of eq.~(\ref{eq:slup}) and the 
label $i$ stands for the order in which the $\cN=n N^n$ different links
are being updated. It is actually not difficult to compute exactly by 
which amount the action is decreasing. For the single-link update we 
find 
\be
S(U^\prime_\mu(x))-S(U_\mu(x))=-2N^{4-n}\|\tilde{U}_\mu(x)\|~\|U^\prime_\mu(x)-
U_\mu(x)\|^2\quad,\label{eq:delS}
\ee
where we made use of eq.~(\ref{eq:slup}). To find the change 
of the action after one sweep, one has to simply add the contributions of 
each single-link update keeping in mind, however, that the value of
$\|\tilde{U}_\mu(x)\|$ depends on which links had been updated before.
In the continuum limit $\|\tilde{U}_\mu(x)\|\rightarrow 2(n-1)$ and
to a good accuracy the action changes from one sweep to the next
by the amount 
\be
S(T(U))-S(U)\sim-4(n-1)N^{4-n}\|T(U)-U\|^2
\equiv-4(n-1)N^{4-n}\sum_{x,\mu}\|U^\prime_\mu(x)-U_\mu(x)\|^2\quad.
\label{eq:dSsw}\ee
For saddle points, which are fixed points of this algorithm, one can 
of course lower the action further and the algorithm is necessarily 
unstable.

For the functional $\hat S$ any solution corresponds to the absolute
minimum $\hat S=0$, and cooling with this functional should hence not
lead to any instability for the saddle points
of the original action, eqs.~(\ref{eq:Swil},\ref{eq:Seps}). 
Unfortunately, it can
be proven that one cannot analytically minimize $\hat S$ as a function
of one of its links. All that is required, however, is that the functional
is lowered if and only if we are not at a fixed point. We will see, as 
our intuition tells us, that this is a sufficient condition for 
convergence of the algorithm (necessity should be obvious). Let us nevertheless
write down the necessary ingredients to minimize the single-link
dependence of $\hat S$, which was already considered by Van der 
Sijs~\cite{sij}. We restrict ourselves in this paper to the 
version of $\hat S$ derived from the Wilson action, as the analysis 
is prohibitively more complicated for improved actions (for $\tilde S$ the
situation is in either case intractable, because of the appearance of
$\|\tilde U_\mu(x)\|$ in the denominator of eq.(\ref{eq:Sus})).
The analogue of eq.~(\ref{eq:Ssin}) becomes 
\be
\hat S(U_\mu(x))=N^{6-n}\RE\Tr\left(U^\dagger_\mu(x)W_\mu(x)-3
\sum_{\alpha=1}^{2(n-1)} (U^\dagger_\mu(x)V^\alpha_\mu(x))^2-
(U^\dagger_\mu(x)\tilde{U}_\mu(x))^2\right)~,\label{eq:hSsin}
\ee
where the index $\alpha$ runs over the $2(n-1)$ different staples of 
eq.~(\ref{eq:Util}), such that $\tilde{U}_\mu(x)=\sum_\alpha V_\mu^\alpha(x)$.
The expression for $W_\mu(x)$ is a complicated combination of planar and
non-planar Wilson loops, which are best exhibited in terms
of unit vectors $\hat a,~\hat b\in\{\pm\hat 1,\cdots,\pm\hat n\}$
\bea
W_\mu(x)&=&2\sum_{\stackrel{a\neq-b}{a,b\neq\pm\mu}}\wplaqone-\wplaqonem+
2\sum_{\stackrel{a\neq-\mu}{b\neq\pm\mu,\pm a}}
\wplaqtwo-\wplaqtwom+\wplaqthree-\wplaqthreem\quad,\nonumber
\eea
or in explicit form, using the convention that $U_a(x)\equiv 
U_{-a}^\dagger(x-\hat a)$:
\bea
& &\hskip-16mm W_\mu(x)={{2\sum}_{}}_{a,b\neq\pm\mu;a\neq-b}\{U_a(x)
U_b(x+\hat a)U_\mu(x+\hat a+\hat b)U_b^\dagger(x+\hat a+\hat\mu)
U_a^\dagger(x+\hat\mu)\nonumber\\ & &\hskip-4mm-
U_a(x)U_\mu(x+\hat a)U_b(x+\hat a+\hat\mu)
U_\mu^\dagger(x+\hat a+\hat b)U_b^\dagger(x+\hat a)
U_\mu(x+\hat a)U_a^\dagger(x+\hat\mu)\}\nonumber\\
&&\hskip-4mm+{{2\sum}_{}}_{b\neq\pm\mu,\pm a;a\neq-\mu}\{
U_b(x)U_\mu(x+\hat b)U_a(x+\hat b+\hat\mu)
U_b^\dagger(x+\hat a+\hat\mu) U_a^\dagger(x+\hat\mu)
\nonumber\\ & &\hskip-4mm-U_b(x)U_\mu(x+\hat b)U_b^\dagger(x+\hat\mu)
U_a(x+\hat\mu)U_b(x+\hat a+\hat\mu)U_a^\dagger(x+\hat b+\hat\mu)
U_b^\dagger(x+\hat\mu)\nonumber\\
&&\hskip-4mm+U_a^\dagger(x-\hat a)U_b(x-\hat a)U_a(x+\hat b-\hat a)
U_\mu(x+\hat b) U_b^\dagger(x+\hat\mu)\nonumber\\ & &\hskip-4mm-
U_b(x)U_a^\dagger(x+\hat b-\hat a)U_b^\dagger(x-\hat a)U_a(x-\hat a)
U_b(x)U_\mu(x+\hat b)U_b^\dagger(x+\hat\mu)\}\quad.
\label{eq:mess}
\eea
The sum in this expression is restricted to those combinations of $\hat a$ 
and $\hat b$ that do not lead to any backtracking in the Wilson loop
defined by the products of the links.
If one tries to minimize $\hat S(U_\mu(x))$ with respect to $U_\mu(x)$
under the constraint $\|U_\mu(x)\|=1$, one finds
\be
\lambda U_\mu(x)=M(U_\mu(x))-W_\mu(x)\quad.
\ee
Note that the lattice should now be at least three sites in each direction. 
The operator $M$, whose $\mu$ and $x$ dependence was suppressed, is
defined for SU(2) through the equation 
\be
M(U_\mu(x))\equiv 6\sum_\alpha\Tr\left(U_\mu(x)V_\mu^\alpha(x)^\dagger\right)
V_\mu^\alpha(x)+2\Tr\left(U_\mu(x)\tilde{U}_\mu^\dagger(x)\right)
\tilde{U}_\mu(x)\quad,\label{eq:Mdef}
\ee
An attempt to solve for $\lambda$, using the condition that $U_\mu(x)$ is
unitary, leads to a complicated eighth order polynomial, and solving it
numerically for each update would be too costly. Instead, we define
our update by the simple equation
\be
U^\prime_\mu(x)={M(U_\mu(x))-W_\mu(x)\over{\|M(U_\mu(x))-W_\mu(x)\|}}\quad,
\label{eq:hSit}
\ee
The new feature is that the r.h.s. still depends on $U_\mu(x)$ and that
we do not obtain a minimum of eq.~(\ref{eq:hSsin}). Nevertheless, this is 
harmless since some algebra will show that our update does lower $\hat S$
and as we argued above, and shall show in some more detail in the next
section, this is sufficient. One finds
\be
\delta\hat S(U_\mu(x))=-N^{6-n}\left(\|M(U_\mu(x))-W_\mu(x)\|~
\|\delta U_\mu(x)\|^2+<\delta U_\mu(x),M(\delta U_\mu(x))>\right),
\label{eq:dShs}
\ee
where, of course, $\delta U_\mu(x)\equiv U_\mu^\prime(x)-U_\mu(x)$,
$\delta\hat S( U_\mu(x))\equiv \hat S(U_\mu^\prime(x))-\hat S(U_\mu(x))$
and
\be
<\delta U_\mu(x),M(\delta U_\mu(x))>=3\sum_\alpha|\Tr(\delta U_\mu^\dagger(x)
V_\mu^\alpha(x))|^2+|Tr(\delta U_\mu^\dagger(x)\tilde{U}_\mu(x))|^2\quad.
\label{eq:rest}
\ee
In the continuum limit one has $\|M(U_\mu(x))-W_\mu(x)\|
\rightarrow 8(n-1)(2n+1)$.
%whereas $\Tr(U_\mu(x)W_\mu^\dagger(x))$ $\rightarrow\sum_\lambda 48N^{-4}
%\Tr(F_{\mu\lambda}^2(x))$.

We note that lattice artefact solutions for $\hat S$
will not be eliminated by eq.~(\ref{eq:hSit}). For example, 
putting one link to minus the identity (while keeping all others equal to the
identity) is a fixed point of eq.~(\ref{eq:hSit}). It is, however, not a fixed
point for eq.~(\ref{eq:slup}). In practise we first use
$S$ cooling to bring the energy (or action) down to values not too much 
above where we expect a saddle point to occur, after which the $\hat S$
cooling will in general bring it to a smooth solution. One can also 
combine the $S$ and $\hat S$ cooling, by replacing $U_\mu(x)$ on the r.h.s.
of eq.~(\ref{eq:hSit}) with $\tilde{U}_\mu(x)/\|\tilde{U}_\mu(x)\|$.
We have not proven that this always lowers $\hat S$ (although it did
in all cases we considered). 

\section{Convergence}

We will study the convergence of the algorithm by considering the 
linear approximation for the mapping $T$ in eq.~(\ref{eq:swee}).
For $S$ (or $S(\eps)$) cooling, $T_i$ will correspond to eq.~(\ref{eq:slup}),
while for $\hat S$ cooling it will correspond to eq.~(\ref{eq:hSit}).
To warm up we first consider the single-link convergence of the latter,
while keeping all links (except $U_\mu(x)$) fixed. For notational convenience
we drop the $\mu$ and $x$ dependence. Let us assume that $U_o$ is a 
fixed point, and $X$ is a small element of the SU(2) Lie-algebra. We now 
consider $U=e^XU_o$ and compute $X^\prime$ to linear order in $X$
from $U^\prime\equiv e^{X^\prime} U_o$. After some algebra one finds
\be
X^\prime=\|M(U_o)-W\|^{-1}\left(6\sum_\alpha\Tr(XY_\alpha^\dagger)Y_\alpha
+2\Tr(X\tilde{Y}^\dagger)\tilde{Y}\right)\quad,\label{eq:scon}
\ee
where $\tilde Y\equiv\sum_\alpha Y_\alpha$ and $Y_\alpha=\IM(U_o^\dagger 
V_\alpha)\equiv\ihalf(U_o^\dagger V_\alpha-V_\alpha^\dagger U_o)$.
In the continuum limit this is easily seen to imply that $\|X^\prime\|/\|X\|=
\cO(N^{-4})$ and we have verified that the  single-link iteration is indeed
extremely efficient. One iteration typically brings $U$ already
very close to the minimum of eq.~(\ref{eq:hSsin}) (although we would like to
emphasize again that this is not really important, as the other links
that contribute to $\hat S$ will be changed during the sweep).

With respect to the SU(2) Lie-algebra basis $i\tau_k/2$ 
($\tau_k$ the Pauli matrices)
eq.~(\ref{eq:scon}) is symmetric, but this is in general no longer true for 
the linear approximation of $T$, which we will denote by $dT$. The latter is 
determined exactly as above by evaluating $U^\prime=T(e^XU_o)$ and 
extracting $X^\prime_\mu(x)$ to linear order in $X$ from 
$\exp(X^\prime_\mu(x))=U^\prime_\mu(x)U^o_\mu(x)^\dagger$, where 
$U_o$ is the fixed point. Numerically this is easily achieved by 
first obtaining the fixed point to a high accuracy, assuming the 
last value of $U$ can be equated to $U_o$, and then looking at 
the response of one complete sweep to $U=e^XU_o$, if necessary doing
extrapolation in $X\rightarrow 0$ to extract the part linear in $X$.
One might be tempted to prove that $T$ is a so-called contraction,
for which it is required that $\|T(V)-T(U)\|\leq\kappa\|V-U\|$, with
$\kappa<1$. The contraction theorem asserts that in such a case 
there is a unique
fixed point of the mapping $T$. At the linear level, near the fixed
point, this would imply that $dT^tdT$ (which is symmetric and positive 
definite) has no eigenvalues bigger or equal to one. We know, however,
that the fixed point is neither globally nor locally unique.
Indeed, for all cases we studied there are plenty of eigenvalues bigger
than one. If $dT$ would have been symmetric, there would have also
been eigenvalues of $dT$ that are bigger (in absolute value) than one,
which would clearly be in contradiction with the fact that the action
is lowered with each update.

As $dT$ is not a symmetric matrix, we can not diagonalize it. Still, it
can over the complex numbers be brought to the so-called Jordan-normal 
form by a non-singular (complex) transformation $A$
\bea
dT=A^{-1}J~A\equiv 
A^{-1}\pmatrix{J_1& &\ominus\cr &\ddots& \cr\ominus & &J_\ell\cr}A\quad,
\eea
where $J_i$ is the smallest Jordan block, of the form
\bea
J_i\equiv\pmatrix{\lambda_i&1& &\ominus\cr &\ddots&\ddots& \cr
 & & \ddots&1\cr\ominus& & &\lambda_i\cr}\quad,
\eea
with $\lambda_i$ the eigenvalues. When complex, they occur in conjugate
blocks. Different blocks in our conventions may have the same $\lambda_i$,
and blocks of dimension 1 correspond to eigenvalues with equal left and
right eigenvectors. As $dT^k=A^{-1}J^kA$, any block with $|\lambda_i|<1$
will approach zero. Blocks with $|\lambda_i|>1$ would grow without bound
and can never be consistent with the fact $\hat S$ is a decreasing 
functional as a function of the sweeps. What remains are
the eigenvalues $|\lambda_i|=1$. They can, and will, occur for the 
following reason. Suppose $X$ is a zero mode for the Hessian of $\hat S$
at the fixed point $U_o$, then $\hat S(e^XU_o)-\hat S(U_o)=\cO(X^4)$.
On the other hand using eq.~(\ref{eq:dShs}),
$\hat S(e^{X^\prime}U_o)-\hat S(e^XU_o)=\cO(\|X^\prime-X\|^2)$, 
which can only be consistent if $X^\prime=X$. Consequently
$X$ is an eigenvector of $dT$ with eigenvalue 1. The converse is also true,
i.e. an eigenvector with eigenvalue 1 is a zero-mode of the Hessian.
Consequently, there are at least as many unit eigenvalues as
there are gauge degrees of freedom. To be precise, there are $3N^n$ minus
the dimension of ${\rm Stab}(U_0)$ unit eigenvalues.
(${\rm Stab}(U_0)$ is the stabilizer of $U_o$, which is
the subgroup of the gauge group that leaves $U_o$ unchanged.  It will
only for very special configurations be non-trivial). It is slightly more
subtle to argue that $|\lambda_i|=1$ is only compatible with eq.~(\ref{eq:hSit})
if $\lambda_i=1$ (one uses the fact that there can be no linear relation
among the eigenvectors, because $A$ is non-singular, which implies that
after each sweep  the amount by which $\hat S$ is lowered is bounded 
away from zero if $\lambda_i\neq 1$). It is more easy to rule out that
$\lambda_i=1$ can occur in blocks $J_i$ of dimension greater than 1. 
In other words, left and right eigenvectors for the eigenvalue equal to 
1 always will coincide. 

Clearly, the fixed point is only unique up to the gauge 
degrees of freedom and the moduli of the fixed point (by moduli we 
mean the gauge invariant parameter space along which $\hat S=0$ and 
consequently along which $S$ is degenerate), and there is a one to 
one relation with the $\lambda_i=1$. These do not influence the 
convergence of the algorithm. But it is important to realize that 
one has no control over which fixed point will be obtained. 
(The gauge in which the fixed point
is obtained will entirely be determined by the starting configuration, as
$T$ commutes with applying a gauge transformation, i.e. $T({}^gU)={}^gT(U)$.)
It can happen that there are zero-modes of the Hessian that are
not associated to a moduli parameter of the fixed point, i.e. $\hat S$
is no longer degenerate to fourth (or higher) order. In that case one needs 
to go beyond the linear approximation of $T$ and convergence is worse than
exponential. In the generic case, however, the rate in which $\hat S$
decreases is determined by the gap $\hat\mu$ in $|\lambda_i-1|$, or after
$k$ sweeps $\hat S$ decreases the next sweep by an amount 
$\delta\hat S\sim\exp(-2k\hat\mu)$. The same argument holds of course for
$S$ cooling to a (local) minimum, replacing the gap $\hat\mu$ 
by its appropriate value ($\mu$) obtained from the mapping $T$ for $S$ cooling.

We can find a relation between the values of these gaps and the minimal 
non-trivial 
curvature of $S$ at the fixed point if we assume that the gap is due 
to a real eigenvalue and its eigenvector coincides with the fluctuation along 
the direction of minimal curvature. The associated zero-point frequency will 
be denoted by $\omega$ (scaled so as to have the appropriate continuum limit). 
At a fixed point $U_o$ one has for 
$X_\mu(x)\equiv i\delta A_\mu^a(x)\tau_a/(2N)$
\be
S(e^XU_o)=S(U_o)+\half N^{-n}\sum_{x\mu a;y\nu b}\delta A^a_\mu(x) 
H^{ab}_{\mu\nu}(x,y)\delta A^b_\nu(y)+\cO(\delta A^3)\quad,
\label{eq:Hess}
\ee
which provides the definition for the Hessian $H^{ab}_{\mu\nu}(x,y)$
as a $3nN^n$ square matrix, whose eigenvalues will scale to the appropriate
continuum limit for $N\rightarrow\infty$. Hence, $\omega^2$ equals the lowest
non-zero eigenvalue ($\lambda_m$) of the Hessian. It can be easily shown that
for any saddle point of $S$, such that $\hat S(U_o)=0$, the functional 
$\hat S$ has the expansion
\be
\hat S(e^XU_o)=\half N^{-n}\sum_{x\mu a;y\nu b}\delta A^a_\mu(x)
\hat H^{ab}_{\mu\nu}(x,y)\delta A^b_\nu(y)+\cO(\delta A^3)~,\quad 
\hat H = 2 H^2~,\label{eq:hHess}
\ee
where by $H^2$ we mean of course the matrix multiplication of the Hessian of S
with itself. In the same shorthand notation (i.e. dropping the indices),
we assume that under the $S$ cooling $H\cdot X=\lambda_m X$ {\bf and} 
$X^\prime\equiv dT\cdot X=(1-\mu) X$, which allows us to calculate the change 
in the action both from eq.~(\ref{eq:Hess}), giving $\delta S=-4
\omega^2\mu(1-\half\mu)N^{2-n}\|X\|^2$ and from eq.~(\ref{eq:dSsw}), 
giving $\delta S=-4(n-1)\mu^2N^{4-n}\|X\|^2$. Equating these results gives
\be
\mu\sim{\omega^2\over (n-1)N^2}\equiv{\lambda_m\over (n-1)N^2}\quad.
\label{eq:Sgap}
\ee
Similarly, using eqs.~(\ref{eq:dShs},\ref{eq:hHess}) one finds for
cooling with the functional $\hat S$ 
\be
\hat\mu\sim{\omega^4\over (n-1)(2n+1)N^4}\equiv{\lambda_m^2\over (n-1)(2n+1)N^4}
\quad.\label{eq:Shgap}
\ee
This is of course in accordance with ones intuition that the algorithm
can slow down considerably if the functional has directions in 
which it is very shallow. Note that our results indicate that slowing 
down of the algorithm goes proportional to $N^{2+n}$ for $S$ cooling and 
proportional to $N^{4+n}$ for $\hat S$ cooling (in both cases a factor 
$N^n$  comes from the number of links to be updated in a single sweep).

Finally, most of the moduli for a continuum solution
are removed by lattice artefacts, although we will discuss in the 
next section that in some cases a few of them can survive on a lattice.
An important example of moduli that are removed by lattice artefacts are
those that are related to the translational invariance of the continuum
theory on a torus. As this breaking is due to the fact that integrals in
the continuum are replaced by sums on the lattice, one can easily argue, 
as in ref.~\cite{gar1},
that the breaking is very mild and goes as $\exp(-\rho N)$ where $\rho$
is the typical scale of the continuum solution in units of the length $L$ of
the torus. For small lattices these translational modes can
cause some trouble, but those lattices are usually anyhow too small 
to extract reliable information about the continuum limit. In ref.~\cite{gar1}
we made judicious use of the dependence of the lattice action on the 
moduli parameters, by using the over-improved action $S(\eps=-1)$ (see
eq.~(\ref{eq:Seps})) so as to ensure that the instanton with the maximal
scale parameter allowed by the finite volume is obtained as a 
local minimum for the action. With $S$ cooling we could then find rather 
accurate lattice solutions, whose scaling to a continuum solution was 
studied in detail. One of the advantages of the $\hat S$ cooling is that 
one can verify if the fixed point has been reached, as $\hat S$ should
approach zero. This can, of course, also be used as a check under $S$
cooling. For example, we performed for an instanton on an $8^3\times 24$
lattice cooling for about $6600$ sweeps, which allowed for a very precise
exponential fit to the decay of the action. The effective value for $\mu$
deduced from this fit is $9.8\times10^{-5}$ (see fig.~3 of ref.~\cite{gar1}).
We find for the best configuration of this four dimensional instanton (see 
table I of ref.~\cite{gar1}) $\hat S(\eps=-1)=4.8\times10^{-4}$ and 
for $\tilde S(\eps=-1)=1.3\times10^{-6}$. The results obtained with $S(\eps=0)$
cooling on this configuration (see the same table), lead to
$\hat S(\eps=0)=2.0\times10^{-6}$, $\tilde S(\eps=0)=8.2\times10^{-9}$, 
and effectively $\mu\sim10^{-6}$ (which can be understood since for $S(\eps=0)$
the $\cO(a^2)$ corrections to the action are canceled, removing to this order
the dependence of the action on the scale parameter. Using the heuristic
estimates of $\mu$ discussed above we find $\mu\sim\omega^2/(3N^2)$. Since
$\omega(\eps\neq 0)=\cO(N^{-1})$, whereas $\omega(\eps=0)=\cO(N^{-2})$, one 
obtains with $N=8$ the right orders of magnitude for $\mu$).
In the remainder we will discuss the new results for $n=3$,
i.e. results for the Yang-Mills energy functional.

\section{Results: Constant solutions}

We will first describe the case of constant curvature solutions, which are
known analytically both on the lattice and in the continuum. We take from 
ref.~\cite{baa2} the following results. If we introduce the so-called twist
tensor $n_{ij}$, which is antisymmetric and takes integer values
(equivalently specified by the magnetic flux $m_i=\half\eps_{ijk}n_{jk}$), the 
following configuration is a solution on a symmetric three torus of length 
$L=1$, or a symmetric lattice with $N^3$ sites
\be
A^o_j(x)=i(-\pi n_{jk}x_k+C_j){\tau_3\over 2}\quad,\quad U^o_i(x)=
\exp(A^o_i(x/N)/N)
\ee
These fields are periodic up to a gauge transformation, to be 
specific~\cite{tho1}
\be
A_j(x+\hat k)=\Omega_k(x)(A_j(x)+\partial_j)\Omega_k^{-1}(x)\quad,\quad
U_j(x+N\hat k)=\Omega_k(x/N)U_j(x)\Omega_k^{-1}((x+\hat j)/N)\quad,
\label{eq:twist}
\ee
where
\be
\Omega_k(x)=\exp(\half i\pi x_j n_{jk}\tau_3)\quad.
\ee
If some of the components of the twist matrix are odd, quantities that
transform in the fundamental representation of SU(2) are anti-periodic,
which can be most easily seen from the Polyakov line
\be
P_j(x)=\half\Tr\{{\rm P}\exp(\int_0^1~ds~A^o_j(x+s\hat j))\Omega_j(x)\}=
\cos(\half C_j-\pi n_{jk}x_k)\quad.\label{eq:Poly}
\ee
The identical result follows on the lattice, after replacing $x$ by $x/N$.
For the continuum Yang-Mills and lattice Wilson actions (energies) we find
\be
S=\pi^2\sum_{ij}n_{ij}^2\quad,\quad S=4N^4\sum_{ij}\sin^2
\left({\pi n_{ij}\over 2 N^2}\right)\quad,\label{eq:Sanal}
\ee
which are constant and independent of $\vec C$. The latter form the gauge
invariant moduli of these solutions, both on a lattice and in the continuum.
If the action is non-zero, i.e. the twist is non-trivial, two of 
the moduli can also be identified with translational invariance, whereas 
translation in the other independent direction (which forms a zero eigenvector
of the twist matrix) leaves the configuration invariant (on the lattice 
there is of course no continuous translation invariance). The stabilizer
of this configuration is the subgroup of constant abelian gauge transformations,
which is one dimensional. On the lattice this means that the Hessian has
$3N^3+2$ exact zero-modes and that $dT$ has the same number of unit eigenvalues
(which we verified numerically). In the continuum we can even compute the 
Hessian exactly~\cite{baa2}. The eigen modes are products of a 
theta function (compare the study in the context of the ``Copenhagen 
vacuum''~\cite{amb}) and a plane wave (purely plane waves for abelian 
fluctuations). One easily shows that the gauge field is (anti) periodic
in the $\vec m$ direction, i.e.
$\vec A(\vec x+\vec m/e)=\tau_3^\Delta\vec A(\vec x)\tau_3^{-\Delta}$,
with $e\equiv|{\rm g.c.d. }(m_i)|$ and $\Delta(\vec m)=0$ or 1, its value 
being determined from (repeated use of) eq.~(\ref{eq:twist}).
The spectrum is now given by 
\be
\lambda_\pm(n,\ell)=2\pi\|\vec m\|(2n+1\pm2)+e^2
((2\ell+\Delta)\pi+\vec C\cdot\vec m/e)^2/\|\vec m\|^2
~,\quad \lambda_o(\vec k)=(2\pi\vec k)^2~,
\ee
The multiplicities are $2|e|$ for $\lambda_\pm$ and 2 for $\lambda_o$.
The parameters $(n,~\ell~,\vec k)$ run over $(\integer,\zahlen,\zahlen^3)$. 
On the lattice $\lambda_o(\vec k)$ will be replaced 
by $4N^2\sum_i\sin^2(\pi k_i/N)$, but the effect of the lattice
on $\lambda_\pm$ is not expected to admit an analytic expression.
In fig.~1a-c we plot the first few eigenvalues of the Hessian for 
resp. $\vec m=(0,0,1)$, $\vec m=(0,0,2)$ and $\vec m=(1,1,1)$, computed 
numerically for $N=4$ and 6 in comparison with the exact continuum result
as a function of $\vec C$, of which only the component along $\vec m$ is 
relevant. Only for $\vec m=(1,1,1)$ is $\Delta$ non-trivial.

A new feature that we believe not to have been observed so far, is 
that there are constant ($\neq0$) curvature solutions on a torus that are 
marginally stable. To be precise, for $\|\vec m\|=1$ and 
$|\vec C\cdot\vec m-\pi|\leq\pi-\sqrt{2\pi}$ (using the $2\pi$ periodicity 
of the spectrum in $\vec C\cdot\vec m$ we can restrict to 
$\vec C\cdot\vec m\in[0,2\pi]$) there are no negative eigenvalues for the 
Hessian. We call this marginal stability, as one can first bring 
$\vec C\cdot\vec m$ outside the above range without changing the energy, and 
then let the configuration decay along one of the two unstable modes. 
For $\|\vec m\|\neq1$ one can easily see that this effect doesn't occur. 

Nevertheless, in all cases considered in fig.~1 the Hessian can have accidental 
zero eigenvalues for particular values of $\vec C$, such that potentially the 
algorithm can slow down dramatically. In table I we compare the gap
$\hat\mu$ (the distance of the largest non-unit eigenvalue of $dT$ to 1) with
the predicted value $\hat\mu=\lambda_m^2/(14 N^4)$, determined from the 
smallest non-zero (absolute) eigenvalue $\lambda_m$ of the Hessian for $S$ 
(see eq.(\ref{eq:Shgap}). The agreement is (perhaps surprisingly) very good. 
For $\vec m=(1,1,1)$ and $N=4$ there is considerable slowing down. What is 
called slowing down is of course dependent on the available computer power. We 
consider something slowed down considerably if (several) 10.000 or more sweeps 
are required to half the value of $\hat S$. In the case of
slowing down, we could generally achieve good results by performing a small
number of $S$ coolings, when the $\hat S$ cooling is being ``stuck'',
sometimes after a (very) small random update on the links.
That $S$ cooling can sometimes help can be argued for the case that
$\lambda_m>0$ as follows. We have seen that to
some degree the $\lambda_m$ eigenvector is also an eigenvector
of $dT$, corresponding to the eigenvalue $\lambda=1-\hat\mu$. This should
be the dominant mode along which the configuration differs from the true
fixed point. All other modes, in particular the unstable ones, should have 
decreased significantly under the $\hat S$ cooling. If we don't cool too long 
with $S$, the unstable mode does not get a chance to grow, whereas the
direction of the stable mode keeps on contracting (faster by a factor 
$14N^2/\lambda_m$). With $\hat S$ cooling one can re-contract the unstable
mode, which might have been grown during the $S$ cooling. We expect that 
the extra computational effort will not exceed by much what one would have 
to spend in trying to accelerate the algorithm, e.g. using 
fast-fourier~\cite{dav,dun} or multi-grid~\cite{hul} methods. 

\vskip5mm
\hskip-8mm\hbox to \hsize{\hfil\vbox{\offinterlineskip
\halign{&\vrule#&\ $#\mystrut$\hfil\ \cr
\noalign{\hrule}
&N&&\quad~\vec m&&~\vec C\cdot\vec m&&\quad\quad\hat S
&&\quad\quad~S&&\quad\quad S_{\rm th}&&\quad\lambda_m&&\quad\quad\hat\mu&&
\quad\quad\hat\mu_{\rm th}&\cr
height 4pt&\omit&&\omit&&\omit&&\omit&&\omit&&\omit&&\omit&&\omit&&\omit&\cr
\noalign{\hrule}
height 4pt&\omit&&\omit&&\omit&&\omit&&\omit&&\omit&&\omit&&\omit&&\omit&\cr
&4&&(1,1,1)&&3.472&&2.90\cdot10^{-4}&&59.02774641&&59.02761860&&
\phm0.051&&6.49\cdot10^{-7}&&7.31\cdot10^{-7}&\cr
&6&&\omit&&2.517&&1.60\cdot10^{-2}&&59.17251733&&59.18005528&&-0.574&&
1.74\cdot10^{-5}&&1.82\cdot10^{-5}&\cr
&8&&\omit&&3.175&&2.65\cdot10^{-6}&&59.20573728&&59.20573660&&
\phm1.906&&\omit&&6.34\cdot10^{-5}&\cr
height 4pt&\omit&&\omit&&\omit&&\omit&&\omit&&\omit&&\omit&&\omit&&\omit&\cr
\noalign{\hrule}
&6&&(0,0,2)&&4.229&&5.24\cdot10^{-7}&&78.75660845&&78.75660838&&
\phm4.084 &&8.58\cdot10^{-4}&&9.19\cdot10^{-4}&\cr
height 4pt&\omit&&\omit&&\omit&&\omit&&\omit&&\omit&&\omit&&\omit&&\omit&\cr
\noalign{\hrule}}}\hfil}
\vskip3mm
{\narrower\narrower{\noindent
Table I: Results for the constant curvature solution on a $N^3$ lattice with 
magnetic flux $\vec m$. $S_{\rm th}$ is the value obtained from the exact 
lattice result given in eq.~(\ref{eq:Sanal}), whereas $\hat\mu_{\rm th}=
\lambda_m^2/(14N^4)$, the predicted value for the distance of the largest 
non-unit eigenvalue of $dT$ to unity. The smallest (in absolute value) 
non-zero eigenvalue for the Hessian of $S$ is indicated by $\lambda_m$. 
This, as well as $\hat\mu$, was obtained from the exact lattice solution 
with the tabulated values of $\vec C\cdot\vec m$ extracted from the data.
}\par}
\vskip5mm

Not only $\hat S$ itself is a good indication for how far one is from
the fixed point. Also $dT$ can, and usually does, have an eigenvalue bigger
than 1 if one is not close enough to the fixed point (the gauge modes 
always correspond to unit eigenvalues), whereas only {\bf at} 
a fixed point will the gauge modes be exact zero-modes for the Hessian.
All this provides us with more than enough criteria to be sure that we have an
accurate lattice solution of the equations of motion. In table I we also
compare the energy of the constant curvature solutions with the exact 
lattice result of eq.~(\ref{eq:Sanal}). The agreement is excellent,
except perhaps for $N=6$ and $m=(1,1,1)$, but in that case we did not
try to cool the configuration down to a very small value of $\hat S$.
Nevertheless, the error in the energy is smaller than one part in 6000. Also
quantities like the Polyakov loops agree very well with the exact lattice
expressions, eq.~(\ref{eq:Poly}).

\section{Results: The sphaleron}

We now discuss the results for the new sphaleron solutions. Some of the 
results are collected in table II. Also here we see that convergence of the
$\hat S$ cooling algorithm can be well predicted in terms of $\lambda_m$.
In figs.~2a-c we show scatter plots for the (complex) eigenvalues of $dT$ for 
the $\vec m=\vec 0$ sphaleron, using a checkerboard updating for fig.~2a
(to allow for vectorization of the algorithm) and a sequential updating for
fig.~2b, both at $N=4$. We see that the checkerboard updating makes the bulk 
of the modes contract much faster (in addition to the gain of computational 
speed due to the vectorization). This remains true on larger lattices. 
In fig.~2c we show for comparison the checkerboard result for $N=8$. 
The eigenvalues that are responsible for the slowing down of the algorithm 
near the fixed point, in particular the eigenvalue closest to 1, will only 
very weakly depend on the order in which one updates the links.
Table II shows that the sphaleron with periodic boundary conditions 
($\vec m=\vec 0$) where $\lambda_m\sim1.3$, has considerable slower cooling than
the sphaleron with twisted boundary conditions, for which $\lambda_m\sim-10.8$.
Nevertheless, we managed to obtain rather accurate results. 
In table II we also list $S$, $\hat S$ and the blocked
Wilson action $S_{2\times2}$, for which the lattice spacing is effectively 
twice as large. Like the instantons of maximal size studied in 
ref.~\cite{gar1,gar2}, the sphaleron is a very smooth solution.
We do not require huge lattices to approach the continuum limit. 
Together with easily obtainable expansions in the lattice spacing $a=1/N$, 
we have a large enough window (available values for $N$) to get accurate 
results. We can extract, as for the instantons (see ref.~\cite{gar1} for 
details), the continuum action by fitting to the formula 
\be
S_{n\times n}=S_0\left(1-\alpha n^2/N^2-(\beta n^4+\gamma n^2+\delta)/N^4
\right)+\cO(n^6/N^6)\quad.
\ee
We find for the twisted continuum sphaleron ($\vec m=(1,1,1)$) an energy
$\cE=34.148(2)$ and for the periodic one ($\vec m=\vec 0$) we find
$\cE=72.605(2)$, with conservative error estimates.
\vskip5mm
\hbox to \hsize{\hfil\vbox{\offinterlineskip
\halign{&\vrule#&\ $#\mystrut$\hfil\ \cr
\noalign{\hrule}
&\pho N&&\quad~\vec m&&\quad~\lambda_m&&\quad\quad\hat\mu&&\quad\quad
\hat\mu_{\rm th}&&\quad~S&&\quad S_{2\times2}&&\quad\quad\hat S&\cr
height 4pt&\omit&&\omit&&\omit&&\omit&&\omit&&\omit&&\omit&&\omit&\cr
\noalign{\hrule}
height 4pt&\omit&&\omit&&\omit&&\omit&&\omit&&\omit&&\omit&&\omit&\cr
&\pho4&&(1,1,1)&&-11.128&&2.73\cdot10^{-2}&&3.46\cdot10^{-2}&&33.24878&&
29.77153&&7.90\cdot10^{-6}&\cr
&\pho6&&\omit&&-10.900&&8.74\cdot10^{-3}&&6.55\cdot10^{-3}&&33.77789&&
32.49357&&2.28\cdot10^{-7}&\cr
&\pho8&&\omit&&-10.817&&\omit&&2.04\cdot10^{-3}&&33.94548&&33.28093&&
1.20\cdot10^{-11}&\cr
&12&&\omit&&\omit&&\omit&&4.03\cdot10^{-4}&&34.05973&&33.78353&&
2.34\cdot10^{-6}&\cr
height 4pt&\omit&&\omit&&\omit&&\omit&&\omit&&\omit&&\omit&&\omit&\cr
\noalign{\hrule}
&\pho4&&(0,0,0)&&\phm\pho1.704&&7.33\cdot10^{-4}&&8.10\cdot10^{-4}&&
68.13887&&52.59651&&2.73\cdot10^{-5}&\cr
&\pho6&&\omit&&\phm\pho1.415&&1.08\cdot10^{-4}&&1.10\cdot10^{-4}&&
70.70174&&64.36296&&2.11\cdot10^{-5}&\cr
&\pho8&&\omit&&\phm\pho1.301&&2.41\cdot10^{-5}&&2.95\cdot10^{-5}&&
71.55060&&68.16790&&6.93\cdot10^{-9}&\cr
&12&&\omit&&\omit&&\omit&&5.83\cdot10^{-6}&&72.14095&&70.70631&&
4.13\cdot10^{-5}&\cr
height 4pt&\omit&&\omit&&\omit&&\omit&&\omit&&\omit&&\omit&&\omit&\cr
\noalign{\hrule}}}\hfil}
\vskip3mm
{\narrower\narrower{\noindent
Table II: Results for the sphalerons on a $N^3$ lattice with twisted and 
periodic boundary conditions. The smallest (in absolute value) non-zero 
eigenvalue for the Hessian of $S$ is indicated by $\lambda_m$. The 
predicted value for the distance of the largest non-unit eigenvalue of 
$dT$ to unity is given by $\hat\mu_{\rm th} =\lambda_m^2/(14N^4)$. We also 
give the value of the Wilson action, its blocked value $S_{2\times 2}$ 
and $\hat S$.
}\par}
\vskip5mm

The evidence that these configurations are sphalerons follows from a study
of the eigenvalues for the Hessian of $S$, whose low-lying values are listed
in table III for $N=4,~6$ and $8$. 
We have verified that there are exactly $3N^3+3$ zero-modes for both 
sphalerons, corresponding to the gauge modes and the three translational 
moduli (for $N>6$ the latter are indistinguishable from the gauge zero-modes, 
for $N=4$ their eigenvalues are not bigger than 0.01). Most importantly we 
have verified that there is precisely one unstable mode. Furthermore, we have 
not been able to find any other saddle point with a lower (non-zero) energy 
than the energies of the present configurations. For all practical purposes 
we thus consider the newly found solutions to be sphalerons. 

We now wish to show that like on $S^3$~\cite{baa1}, these sphalerons are
at the top of a tunnelling path. For the case of twisted boundary conditions
this is the easiest to demonstrate, since the instanton with a continuum
action of $4\pi^2$, which was extensively studied on the lattice
in ref.~\cite{gar3}, has only the trivial translational and some
discrete moduli. This means we only have to demonstrate that the 
top of the instanton path is identical to the sphaleron. There are two reasons
why the top of an instanton path might not correspond exactly to a sphaleron.
First, the lattice has a finite resolution and second, the extent in the time 
direction is finite. In particular the finite resolution causes problems
because one needs to extrapolate along the tunnelling path the value
of the energy and of $\hat S$. In particular the latter has a sharp dip
near the top. A very large value of $N$ would be required to obtain
a sufficient resolution. The effect of the finite extent in time
is expected to be small since the instanton path approaches the vacua 
exponentially in time~\cite{gar3}.  In fig.~3 we plot $\tilde S
\sim\hat S/64$ for each time-slice along the tunnelling path, 
cmp. ref.~\cite{gar2}.

\vskip5mm
\hbox{\hskip1.5cm$\vec m=(1,1,1)$ sphaleron\hskip2.5cm$\vec m=\vec 0$ 
sphaleron\hskip3cm index 2}
\hbox to\hsize{\hfil\vbox{\offinterlineskip\halign
{&\vrule#&\ $#\mystrut$\hfil\ \cr \noalign{\hrule}
&\quad N=4&&\quad N=6&&\quad N=8&&\quad&&\quad N=4&&\quad N=6&&\quad N=8&&
\quad&&\quad N=8&\cr
height 4pt&\omit&&\omit&&\omit&&\omit&&\omit&&\omit&&\omit&&\omit&&\omit&\cr
\noalign{\hrule}
height 4pt&\omit&&\omit&&\omit&&\omit&&\omit&&\omit&&\omit&&\omit&&\omit&\cr
&-11.1280&&-10.9003&&-10.8174&&\omit&&-13.1960&&-13.3935&&-13.4432&&\omit&&
-13.1170&\cr
&\phm\pho0.0074&&\phm\pho0.0001&&\phm\pho0.0000 &&\omit&&\phm\pho0.0073&&
\phm\pho0.0006&&\phm\pho0.0000 &&\omit&&-\pho0.7466&\cr
&\phm\pho0.0091&&\phm\pho0.0001&&\phm\pho0.0000 &&\omit&&\phm\pho0.0124&&
\phm\pho0.0010&&\phm\pho0.0000 &&\omit&&\phm\pho0.0000&\cr
&\phm\pho0.0101&&\phm\pho0.0001&&\phm\pho0.0000 &&\omit&&\phm\pho0.0139&&
\phm\pho0.0011&&\phm\pho0.0000 &&\omit&&\phm\pho0.0000&\cr
&\phm13.1667&&\phm13.3371&&\phm13.3934 &&\omit&&\phm\pho1.7041&&
\phm\pho1.4151&&\phm\pho1.3009 &&\omit&&\phm\pho0.0000&\cr
&\phm13.1667&&\phm13.3371&&\phm13.3934 &&\omit&&\phm\pho1.7070&&
\phm\pho1.4303&&\phm\pho1.3012 &&\omit&&\phm\pho3.2618&\cr
&\phm13.1667&&\phm13.3372&&\phm13.3934 &&\omit&&\phm\pho7.9677&&
\phm\pho8.0984&&\phm\pho8.1260 &&\omit&&\phm\pho5.8733&\cr
&\phm16.0475&&\phm16.8663&&\phm17.1597 &&\omit&&\phm\pho7.9682&&
\phm\pho8.1008&&\phm\pho8.1261 &&\omit&&\phm\pho5.8807&\cr
&\phm16.0478&&\phm16.8663&&\phm17.1597 &&\omit&&\phm\pho7.9684&&
\phm\pho8.1279&&\phm\pho8.1266 &&\omit&&\phm\pho6.1738&\cr
&\phm16.0484&&\phm16.8663&&\phm17.1597 &&\omit&&\phm\pho8.9984&&
\phm10.2977&&\phm10.6604 &&\omit&&\phm\pho8.5986&\cr
&\phm21.4686&&\phm22.9444&&\phm23.4481 &&\omit&&\phm\pho9.0189&&
\phm10.3306&&\phm10.6611 &&\omit&&\phm\pho8.6069&\cr
&\phm21.4747&&\phm22.9444&&\phm23.4481 &&\omit&&\phm\pho9.9919&&
\phm10.8287&&\phm11.0701 &&\omit&&\phm15.0281&\cr
&\phm21.8119&&\phm23.6594&&\phm24.2198 &&\omit&&\phm10.0256&&
\phm10.8312&&\phm11.0701 &&\omit&&\phm16.7321&\cr
&\phm21.8260&&\phm23.6594&&\phm24.2198 &&\omit&&\phm10.0354&&
\phm10.8594&&\phm11.0707 &&\omit&&\phm19.7489&\cr
&\phm25.0998&&\phm26.1760&&\phm26.4962 &&\omit&&\phm24.6675&&
\phm26.4047&&\phm26.9618 &&\omit&&\phm20.6039&\cr
&\phm25.1050&&\phm26.1760&&\phm26.4962 &&\omit&&\phm24.6703&&
\phm26.4205&&\phm26.9621 &&\omit&&\phm21.3605&\cr
%&\phm25.1146&&\phm26.1761&&\phm26.4962 &&\omit&&\phm25.0288&&
%\phm26.6866&&\phm27.3008 &&\omit&&\phm22.2329&\cr
height 4pt&\omit&&\omit&&\omit&&\omit&&\omit&&\omit&&\omit&&\omit&&\omit&\cr
\noalign{\hrule}}}\hfil}
\vskip3mm
{\narrower\narrower{\noindent
Table III: The lowest lying eigenvalues for the Hessian of $S$ for the 
twisted and periodic sphaleron on a $N^3$ lattice for $N=4,~6$ and 8. 
Also is given the spectrum for the periodic solution with two unstable 
modes (index 2), close in energy to the sphaleron, discussed in the text 
(see also table IV).
}\par}
\vskip3mm

We take the configuration at the top of the instanton path, where
$\tilde S$ is minimal, and
cool it with $\hat S$. This way we easily obtain the sphaleron solution.
As $\hat S$ decreases monotonically under the $\hat S$ cooling, and
as $\hat S$ is already small at the top of the instanton path,
we expect that the instanton path goes through a sphaleron. Further
evidence is provided by calculating the quantity 
\be
D_N(t)=4\|U_t-U_s\|^2/N\sim\int d_3x (\delta A^a_i(\vec x))^2\quad, 
\label{eq:dist}
\ee
where the latter expression is obtained in the continuum limit.
$U_t$ is the configuration at the time $t$ along the instanton path and $U_s$
is the sphaleron configuration, obtained after cooling from the configuration
$U_t$ (for this, $t$ should be chosen not too far away from the top of the 
barrier, as otherwise cooling will bring the configuration to the vacuum).
As the cooling algorithm commutes with gauge transformations, this gives
a gauge invariant measure for the distance of the two configurations. Indeed, 
computing $\|U_t-U_{t^\prime}\|$ gives huge values, as the gauge between
different time slices will differ randomly. For the configurations of fig.~3
we find $D_8(0)=0.0246$, $D_8(1/8)=1.162$, $D_8(-1/8)=1.5713$ and
$D_{12}(0)=0.0340$. We have indicated also the values at either side of the 
top of the barrier for $N=8$ to show that, not being exactly at the top
can have a sizable effect on $D_N(0)$ (and on $\tilde S(t=0)$). We believe
the numerical evidence is sufficiently convincing to conclude that the 
instanton path does go through a sphaleron. In fig.~4 we give for $N=12$ the 
two dimensional cross sections of $\cE_B(x,y,z)$ at the twelve lattice 
values of $x$. Because of the cubic symmetry of the sphaleron, the other 
cross-sections have identical profiles (modulo translations). 

For the case of periodic boundary conditions the instantons have
additional moduli. 
We did provide evidence in ref.~\cite{gar1} that for $T\rightarrow\infty$ 
the moduli are described by the scale, translational and vacuum parameters.
The latter being specified by the Polyakov loops $P_i(t=\infty)=
-P_i(t=-\infty)$ (we have not yet proven or disproven if this relation 
at $t=\pm\infty$ can be relaxed). 
In particular we probed the dependence of the tunnelling 
path on $P_3\equiv P_3(t=-\infty)$, keeping the other two equal to 
unity. In practise we can of course not reach $|t|=\infty$, and this 
is a problem for $P_3=\pm1$. In that case one can show that the approach to 
the vacuum for large $t$ is no longer exponential due to the quartic 
behaviour of the potential in the zero-momentum sector. We demonstrate
the sensitivity of the instanton path on the extent in the time direction
for $N=8$ and $N_t=24,~48$ in fig.~5. We see that for large $N_t$ the tails 
in the electric energy~\cite{gar1,gar2} are lowered considerably, while the 
magnetic energy at the top of the barrier (not visible at the scale of fig.~5)
is lowered from $S(\eps=-1)=74.041$ at $N_t=24$ to 73.465 at $N_t=48$, 
actually below the value of 73.923 obtained for $P_3=0.5$ in
ref.~\cite{gar2} (for which we did not observe such a sensitivity in 
$N_t$). It means that there is near degeneracy in energy at the top of the 
barrier as a function of $P_3$. The variation of the energy as a function 
of $P_3$ can be so small, because the energy functional near the sphaleron 
has two rather shallow directions (associated to the two eigen modes with 
$\lambda_m\sim1.3$). To investigate if our sphaleron indeed corresponds to the 
top of a tunnelling path, presumably the one with $P_3=1$, we took the 
configuration at the top of the energy barrier for the values of $P_3$ that 
were considered in ref.~\cite{gar2} and applied $\hat S$ cooling for 8000 
sweeps on each, such that the energy and $\hat S$ stabilized, and further 
cooling would make the configuration only move (very slowly) along the slowest 
mode in the cooling algorithm. We discovered a very useful set of parameters
to distinguish the various configurations thus obtained. It turns out that
the energy profile is almost identical for each configuration, that the 
profiles for the Polyakov lines only vary weakly, but that the vector 
\be
\vec b\equiv\left(\Tr(B_1^2),\Tr(B_2^2),\Tr(B_3^2)\right)/
\sqrt{\third\sum_i\Tr(B_i^2)^2}\quad,\label{eq:orpa}
\ee
averaged over the lattice ($<\vec b>$), clearly differentiates the various 
configurations. The sphaleron solution distinguishes itself by a symmetric 
value $\vec b=(1,1,1)$ at each point independently. In all other cases the 
variance of the parameter $\vec b$ is non-zero. To a good degree
$<b_1>=<b_2>$, which is of course a direct consequence of the symmetry of 
the lattice and of the fact that $P_1=P_2$. We also compute $D^\prime_8$,
which is defined as in eq.(\ref{eq:dist}) but with $U_t$ the configuration
obtained after the $\hat S$ cooling. For this it is important to note
that the instantons in table IV were all generated from the same random start
at $P_3=1$ (or $C_3=0$; see ref.~\cite{gar2}). This means that each is in 
the same gauge and $D^\prime_8$ is a good gauge invariant measure for the 
distance of the various cooled configurations to the sphaleron.
In table IV we collect the results thus obtained, and
see that after 8000 cooling sweeps $\hat S$ is dramatically lowered 
also for $P_3=0$. If we compute the Hessian for this configuration we find 
a Morse index of two, i.e. there are two unstable modes 
(see table III for the low-lying eigenvalues). 
Note that $S$ for this configuration is only slightly higher than 
for the sphaleron, and that $\lambda_m=-0.74655$ will cause an even slower 
cooling than for the sphaleron. We have performed 8000 extra $\hat S$ cooling 
sweeps before computing the Hessian, but the quality of its zero modes were 
sufficiently good to trust the cooled configuration for $P_3=0$ to be close 
enough to a solution. We notice that except near $P_3=1$, the 
value of $<\vec b>$ at the top of the barrier (denoted by $<\vec b_t>$ in 
table IV) is changed after 8000 $\hat S$ cooling sweeps. We conclude that the 
index 2 solution is not at the top of any of the instanton paths we 
considered here.

\vskip5mm
\hbox to \hsize{\hfil\vbox{\offinterlineskip
\halign{&\vrule#&\ $#\mystrut$\hfil\ \cr
\noalign{\hrule}
&\pho C_3&&\quad\quad<\vec b_t>&&\quad\quad<\vec b>&&\quad~ S&&
\quad S_{2\times 2}&&\quad\quad\hat S&&D^\prime_8&\cr
height 4pt&\omit&&\omit&&\omit&&\omit&&\omit&&\omit&&\omit&\cr
\noalign{\hrule}
height 4pt&\omit&&\omit&&\omit&&\omit&&\omit&&\omit&&\omit&\cr
&\pho0&&(1.01,1.01,0.99)&&(1.01,1.00,0.99)&&71.5507&&68.1680&&
2.08\cdot10^{-4}&&0.000&\cr
&\pho\pi/5&&(1.02,1.02,0.96)&&(1.03,1.02,0.95)&&71.5547&&68.1732&&
9.77\cdot10^{-3}&&0.056&\cr
&2\pi/5&&(1.05,1.05,0.89)&&(1.07,1.06,0.85)&&71.5913&&68.2205&&
7.53\cdot10^{-2}&&0.100&\cr
&2\pi/3&&(1.10,1.10,0.77)&&(1.14,1.13,0.65)&&71.7569&&68.4322&&
1.58\cdot10^{-1}&&0.560&\cr
&4\pi/5&&(1.11,1.11,0.73)&&(1.15,1.15,0.60)&&71.8301&&68.5250&&
1.18\cdot10^{-1}&&0.894&\cr
&\pho\pi&&(1.12,1.12,0.69)&&(1.17,1.16,0.53)&&71.9318&&68.6537&&
2.60\cdot10^{-4}&&2.186&\cr
height 4pt&\omit&&\omit&&\omit&&\omit&&\omit&&\omit&&\omit&\cr
\noalign{\hrule}}}\hfil}
\vskip3mm
{\narrower\narrower{\noindent
Table IV: Results for $N=8$ associated to the top of the instanton path 
(obtained with $S(\eps=-1)$ cooling, see ref.~\cite{gar2}) that interpolates 
from the vacuum specified by $P_1=P_2=1$ and $P_3=\cos(C_3/2)$ at 
$t=-\infty$, to the vacuum specified by $P_1=P_2=-1$ and $P_3=-\cos(C_3/2)$ 
at $t=\infty$. The vector $\vec b$ is the parameter defined in 
eq.~(\ref{eq:orpa}). From the third column on are tabulated the results 
after 8000 $\hat S$ cooling sweeps. $D^\prime_8$ is defined as in 
eq.~(\ref{eq:dist}), with $U_t$ the cooled configuration. Both from the first 
and last case one finds solutions of the equations of motion, resp. the 
sphaleron and the index 2 saddle point. See table III for their Hessians.
}\par}
\vskip5mm

Finally, we calculated $D_N(t)$ from the configuration in table IV with
$C_3=0$ ($P_3=1$) after further $S(\eps=-1)$ cooling on the instanton, as
represented in fig.~5 (at $N_t=24$). We find $D_8(0)=0.0496$, $D_8(-1/8)=2.212$,
$D_8(1/8)=3.369$ and $D_{12}(0)=0.0368$ (not represented in fig.~5).
Taking into account the time-tail effects discussed before and
the fact that the instanton path was generated with $S(\eps=-1)$ cooling, 
whereas $\hat S$ cooling was performed at $\eps=1$, the quality of these 
results are sufficient to conclude that the sphaleron lies on the instanton 
path that has $P_i(t=-\infty)=-P_i(t=\infty) =\pm1$. In fig.~6 we give 
the results for the energy distribution on a $12^3$ lattice.

\section{Conclusions}

The main motivation for this paper was to find the sphaleron solution for
pure SU(2) gauge theories in a finite volume and to show that it is at
the top of the barrier that separates two classical minima, 
connected by an instanton solution with minimal action. We studied both
the case of twisted and periodic boundary conditions. For the latter the 
situation was rather subtle because of the many additional moduli in the 
instanton parameter space. Furthermore, at the sphaleron there are two very 
flat directions for the energy functional, and we found a saddle point with
two unstable modes not more than $\delta\cE=0.5$ above the sphaleron.
The sphaleron solution
for the periodic case has some very special properties. Like for the twisted
case~\cite{gar3} only low fourier components seem to dominate, but 
especially the property that locally $Tr(B_i)^2$ is independent of $i$
is intriguing. There is good hope that in the future this problem might be
tractable analytically. In any case it presents an interesting challenge.

To achieve all these results we built on the idea that one can find 
unstable solutions to the equations of motion by minimizing the functional
obtained by squaring the gradient of the field equations~\cite{dun,sij}.
What we have added is a specific algorithm for SU(2) to minimize this 
functional in a deterministic way. This allowed us to analyse in 
detail the convergence of the algorithm, also providing results for the 
standard cooling algorithm~\cite{ber}. It is likely that the algorithm
can be extended to include a scalar sector so that we can also study 
the sphaleron in the standard model~\cite{kli}. In any case one can 
always try to minimize the functional by doing random updates.
One can easily think of many more applications worthwhile pursuing with 
these methods.

\section{Acknowledgements}

We have benefited from discussions with Jim Hetrick, Jan Smit, Arjan van der 
Sijs and Jeroen Vink. The eigenvalues of the linearized algorithm and Hessian 
were determined using the EISPACK~\cite{eisp} routines. We thank Willem Vermin 
for providing the source codes. Figs.~4 and 6 were produced using 
Mathematica~\cite{math}, its programs with the raw data for the energies are 
attached to the electronic hep-lat bulletin board version of this paper. 
This work was supported in part by grants from ``Stichting voor Fundamenteel 
Onderzoek der Materie (FOM)'' and ``Stichting Nationale Computer 
Faciliteiten (NCF)'' for use of the CRAY Y-MP and C98/4256 at SARA. 
M.G.P. was supported by a Human Capital and Mobility EC fellowship.
\eject

\eject
\begin{center}
\Large{\bf Figure captions}
\end{center}
\vskip1cm
{\narrower\narrower{\noindent
Figure 1: The low-lying eigenvalues of the Hessians for the constant 
curvature solutions specified by (a) $\vec m=(0,0,1)$, (b) $\vec m=(0,0,2)$
and (c) $\vec m=(1,1,1)$ at $N=4$ (dotted lines), $N=6$ (dashed lines) and
$N=\infty$ (i.e. the continuum). Each level is two-fold degenerate
except for $\vec m=(0,0,2)$ and the levels indicated by the arrow in (a),
which are all four-fold degenerate.
}\par}
\vskip1cm
{\narrower\narrower{\noindent
Figure 2: The scatter plots for the complex eigenvalues of $dT$ for the
periodic sphaleron at $N=4$ (a,b) and $N=8$ (c), for the latter we also show
the blown-up region near $\lambda=1$. Figs.~(a,c) were generated
with a checkerboard updating, whereas fig.~(b) was generated with sequential
updating.
}\par}
\vskip1cm
{\narrower\narrower{\noindent
Figure 3: The value of $\tilde S(t)\sim\hat S(t)/64$ (at $\eps=1$) for the 
twisted instanton~\cite{gar3} with continuum action $4\pi^2$ on a 
$8^3\times 24$ (triangles) and a $12^3\times36$ (squares) lattice. 
The strong dip in $\tilde S(t)$ at $t=0$ indicates stationarity 
of the energy functional due to the tunneling through a sphaleron.
}\par}
\vskip1cm
{\narrower\narrower{\noindent
Figure 4: The profile for the magnetic energies $\cE_B$ for the twisted
sphaleron on a $12^3$ lattice at fixed values of $x$ as a function of
$y$ and $z$. The lattice is represented by the grid-lines (all coordinates
are interchangeable). The vertical axes runs from 0 to 110, in units of the 
inverse physical length of the box.
}\par}
\vskip1cm
{\narrower\narrower{\noindent
Figure 5: The electric and magnetic energies obtained after cooling with
$S(\eps=-1)$ for a periodic $8^3$ lattice in the space directions
and for boundary conditions fixed in the time direction of $N_t$ sites
to $P_i=-1$ at one end
and $P_i=1$ at the other end, for $N_t=24$ (squares for $\cE_E$ and crosses
for $\cE_B$) and $N_t=48$ (triangles for $\cE_E$ and stars for $\cE_B$). 
The time-tails are plotted at a blown-up scale to show the electric 
tails~\cite{gar1} that distort the instanton solution at small $N_t$.
}\par}
\vskip1cm
{\narrower\narrower{\noindent
Figure 6: The profile for the magnetic energies $\cE_B$ for the periodic
sphaleron on a $12^3$ lattice at fixed values of $x$ as a function of
$y$ and $z$. The lattice is represented by the grid-lines (all coordinates
are interchangeable). The vertical axes runs from 0 to 110, in units of the 
inverse physical length of the box.
}\par}

\begin{thebibliography}{99}
\bibitem{kli}
F. R. Klinkhamer and M. Manton, Phys. Rev. \un{D30} (1984) 2212.
\bibitem{tau}
C. Taubes, Comm. Math. Phys. \un{86} (1982) 257, 299;\\
C. Taubes, in: Progress in gauge filed theory, eds. G. 't Hooft
et. al., Plenum Press, New York, 1984, p.563.
\bibitem{baa1}
P. van Baal and N. D. Hari Dass, Nucl. Phys. \un{B385} (1992) 185.
\bibitem{gar1}
M. Garc\'{\i}a P\'erez, A. Gonz\'alez-Arroyo, J. Snippe and P. van Baal,
Nucl. Phys. \un{B413} (1994) 535.
\bibitem{gar2}
M. Garc\'{\i}a P\'erez, A. Gonz\'alez-Arroyo, J. Snippe and P. van Baal,
On the top of the energy barrier, Leiden preprint INLO-PUB-14/93,
to appear in the proceedings of Lattice '93, Nucl.Phys. B(Proc.Suppl.) (1994).
\bibitem{gar3}
M. Garc\'{\i}a P\'erez, A. Gonz\'alez-Arroyo and B. S\"oderberg, Phys.
 Lett. \un{B235} (1990) 117; M. Garc\'{\i}a P\'erez and A. Gonz\'alez-Arroyo,
J. Phys. \un{A26} (1993) 2667.
\bibitem{dun}
A. Duncan and R.D. Mawhinney, Nucl. Phys. B(Proc.Suppl.)26 (1992) 444;
Phys. Lett. \un{B282} (1992) 423.
\bibitem{wil}
K. Wilson, Phys. Rev. \un{D10} (1974) 2445.
\bibitem{sij} A.J. van der Sijs, Nucl. Phys. B(Proc.Suppl.)30 (1993) 893;
Phys. Lett. \un{B294} (1992) 391.
\bibitem{ber}
B. Berg, Phys. Lett. \un{104B} (1981) 475;\\
J. Hoek, M. Teper and J. Waterhouse, Nucl. Phys. \un{B288} (1987) 589.
\bibitem{baa2}
P. van Baal,  Comm. Math. Phys. \un{94} (1984) 397. 
\bibitem{tho1}
G. 't Hooft, Nucl. Phys. \un{B153} (1979) 141.
\bibitem{amb}
J. Ambj\o rn and P. Olesen, Nucl. Phys. \un{B170[FS1]} (1980) 60;\\
J. Ambj\o rn, B. Felsager and P. Olesen, Nucl. Phys. \un{B175} (1980) 349.
\bibitem{dav}
C.T.H. Davies, et al, Phys. Rev. \un{D37} (1988) 1581.
\bibitem{hul}
A. Hulsebos, M.L. Laursen and J. Smit, Phys. Lett. \un{B291} (1992) 431.
\bibitem{eisp} B.T. Smith, et al, Matrix Eigensystem
Routines - EISPACK Guide, second edition (Springer, New York, 1976);
B.S. Garbow, et al, Matrix Eigensystem Routines - EISPACK Guide Extension
(Springer, New York, 1977).
\bibitem{math}
S. Wolfram, et. al., Mathematica (Addison-Wesley, New York, 1991).
\end{thebibliography}
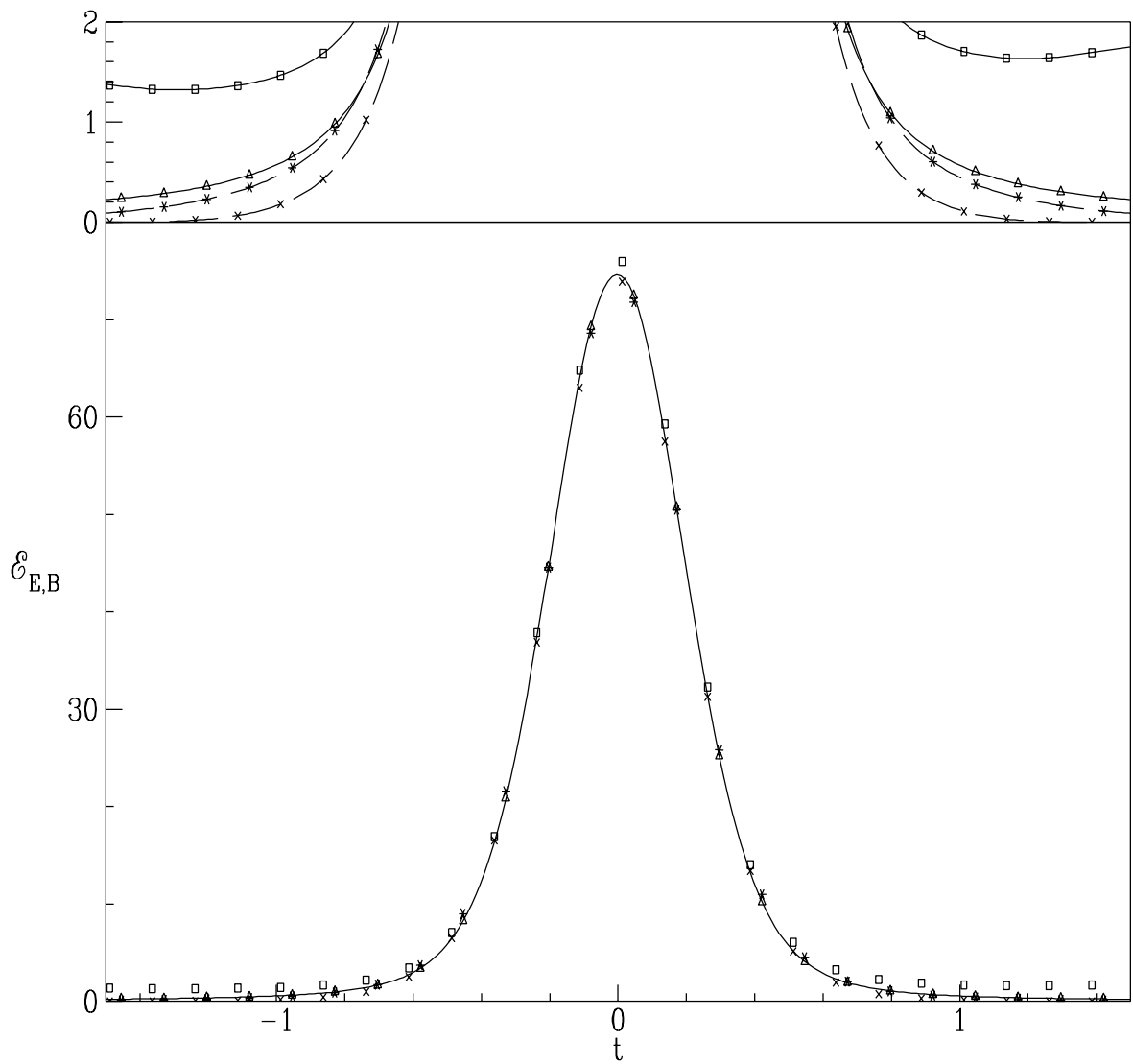
\end{document}